\documentclass[a4paper]{jpconf}
\usepackage{graphicx}

\newcommand{\ycz}{YbCo$_2$Zn$_{20}$}
\newcommand{\bq}{\textit{\textbf{Q}}}

\begin{document}
\title{Effect of magnetic field in heavy-fermion compound YbCo$_2$Zn$_{20}$}

\author{Koji Kaneko$^1$, Shingo Yoshiuchi$^2$, Tetsuya Takeuchi$^3$, Fuminori Honda$^2$, Rikio Settai$^2$ and Yoshichika \={O}nuki$^2$}

\address{$^1$ Quantum Beam Science Directorate, Japan Atomic Energy Agency, Tokai, Naka Ibaraki 319-1195, Japan}
\address{$^2$ Graduate School of Science, Osaka University, Toyonaka 560-0043, Japan}
\address{$^3$ Low Temperature Center, Osaka University, Toyonaka 560-0043, Japan}

\ead{kaneko.koji@gmail.com}

\begin{abstract}
Inelastic neutron scattering experiments on poly crystalline sample of heavy-fermion compound \ycz were carried out in order to obtain microscopic insights on the ground state and its magnetic field response.
At zero field at 300\,mK, inelastic response consists of two features: quasielastic scattering and a sharp peak at 0.6\,meV.
With increasing temperature, a broad peak comes up around 2.1\,meV, 
whereas quasielastic response gets broader and the peak at 0.6\,meV becomes unclear.
By applying magnetic field, the quasielastic response exhibits significant broadening above 1\,T, and the peak at 0.6\,meV is obscure under fields.
The peaks in inelastic spectra and its temperature variation can be ascribed to the suggested crystal-field model of ${{\Gamma}_6}$ - ${{\Gamma}_8}$ - ${{\Gamma}_7}$ with the overall splitting of less than 3\,meV.
The observed quasielastic response and its rapid broadening with magnetic field indicates that the heavy-electron state arises from the ground state doublets, and are strongly suppressed by external field in {\ycz}, 
\end{abstract}

\section{Introduction}
The new family of ternary compounds $RT_2X_{20}$ ($R$=rare earth or actinide, $T$ = transition metal, $X$ = Al, Zn) with the cubic CeCr$_2$Al$_{20}$ structure 
get surge of interest due to its wide variety of physical properties including superconductivity\cite{Tor, Oni}.
The characteristic structure with an $R$ ion incorporated in the nearly spherical cage formed by 16 $X$ ions, gives two essential aspects on physical properties; 
one is strong hybridisation between $f$ and conduction electrons due to a large coordination number around the $R$ ion, 
and the other is highly quasi-degenerated crystal electric field levels. 
Moreover, a large distance between $R$ ions of ${\sim}$6\,{\AA} leads to weaker interactions, and allows various attractive ground states to be realised in this system. 
Among this family, it should be noted that all of ytterbium members Yb$T_2$Zn$_{20}$ ($T$= Fe, Co, Ru, Rh, Os, Ir) exhibit heavy-fermion behaviour with the large electronic specific coefficient of  ${\gamma}\,>$\,400\,mJ/mol K$^2$[1-9]. 
In particular, YbCo$_2$Zn$_{20}$ shows extremely large value of ${\gamma}{\sim}$8\,J/mol K$^{2}$, one of the heaviest fermion compounds so far.  
The $A$ coefficient of electrical resistivity is correspondingly large, and the Kondo temperature is estimated to be less than 3\,K in this compound\cite{Tor,Ohy, Tak2}.
Based on these observation, YbCo$_2$Zn$_{20}$ is considered to lie at the proximity of a quantum critical point (QCP).

In fact, recent study revealed the appearance of ordered phases by applying small pressure and/or external magnetic field in YbCo$_2$Zn$_{20}$\cite{Sai, Tak2, Shi}. 
In the case of Yb based compounds, the application of hydrostatic pressure tunes the system from a non-magnetic ground state to a magnetic ordered one through the quantum critical point. 
Saiga {\textit{et al.}} revealed  pressure induced antiferromagnetic order in YbCo$_2$Zn$_{20}$ at around ${\sim}$1\,GPa,
where deviation from the Fermi liquid behaviour was observed\cite{Sai,Mat2}.
The application of a magnetic field leads to drastic effects.
Metamagnetic transitions were found at low field of $B_{\rm M}{\sim}$0.6\,T, where non-Fermi liquid behaviour was observed as well\cite{Ohy, Tak2}. 
Further increasing the field, the $A$ coefficient of electrical resistivity in YbCo$_2$Zn$_{20}$ exhibits rapid decrease, which suggests that heavy quasiparticles are strongly suppressed by external magnetic field\cite{Sai,Mat,Ohy,Tak2}. 
Besides, a field-induced ordered phase, of which origin is still under investigation, occurs when a field of more than 6\,T is applied along the [1\,1\,1] direction\cite{Tak2,Shi}.
These facts suggest that {\ycz} is located very close to a QCP, and can be one of the model compounds in Yb-system to study dynamical properties around QCP.
In order to get microscopic insights on $f$-electron state in YbCo$_2$Zn$_{20}$,inelastic neutron scattering experiments down to low temperature and under magnetic fields were carried out.

\section{Experiment}
Polycrystalline samples of {\ycz} were grown by the Zn self-flux method.
Details of the sample preparation is published elsewhere\cite{Yos, Tak1, Ohy}.
Neutron inelastic scattering experiments were performed on the cold triple-axis spectrometer LTAS installed at the C2-1 port in the guide hall of the research reactor JRR-3 in Tokai, Japan.
A combination of vertical and horizontal focusing PG monochrometer and analyser, respectively were employed.
The collimation of guide-80'-120'(radial)-open with $k_f$=1.3\,{\AA} gives the energy resolution of about 140\,${\mu}$eV.
The sample was attached to the $^3$He-$^4$He dilution refrigerator, and inserted into the 5\,T vertical field superconducting magnet.
\begin{figure}[t]
	\begin{center}
		\includegraphics[width=9.5cm]{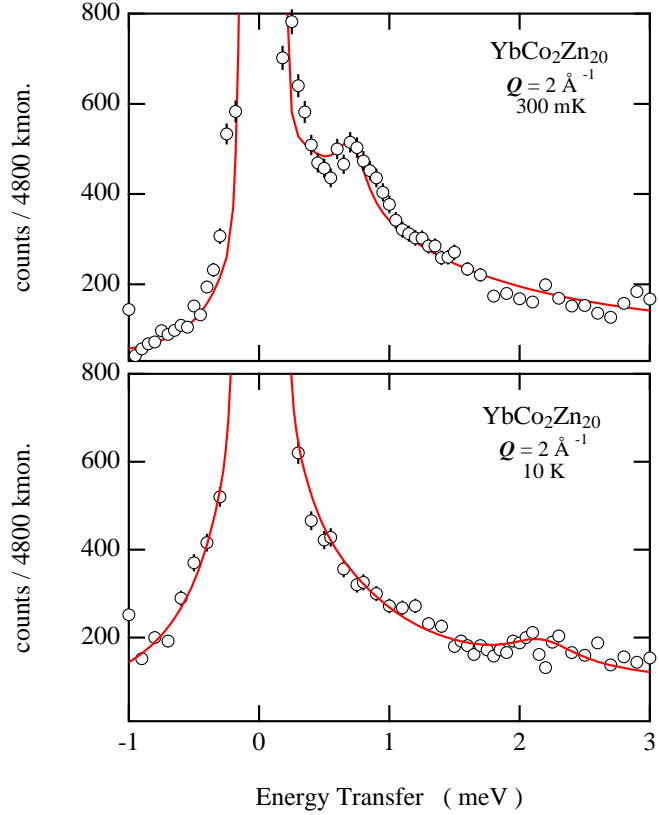}%
		\caption{\label{Fig:1}Inelastic spectra measured on polycrystalline sample of {\ycz} at (a) 300\,mK and (b) 10\,K with $k_f$=1.3\,{\AA}$^{-1}$.\vspace{-5mm}}
	\end{center}
\end{figure}

\section{Results and Discussion}

%Figure 1 shows inelastic scattering spectra of \ycz at \bq=2\,{\AA}$^{-1}$ measured at 50\,mK and 10\,K.
%At the base temperature of 50\,mK, strong quasielastic response was observed.
Figure 1 shows inelastic scattering spectra of {\ycz} at \bq=2\,{\AA}$^{-1}$ up to 3\,meV measured at (a)\,300\,mK and (b)\,10\,K.
At 300\,mK, strong quasielastic response which extends above present measured range was observed in {\ycz}.
Besides, a distinct inelastic peak was superposed onto the quasielastic response at around 0.6\,meV.
With increasing temperature up to 10\,K, inelastic spectra of {\ycz} undergoes substantial change.
It should be noted that an additional excitation appears around 2.1\,meV.
On the other hand, the excitation at 0.6\,meV becomes unclear, and the quasielastic part gets broader.

Figure 2 displays magnetic field dependence of inelastic scattering spectra of {\ycz} measured at 50\,mK under several fields up to 5\,T.
Note that inelastic scattering spectra of {\ycz} exhibit strong magnetic-field variation.
At low magnetic field up to 1\,T, the identical spectra to that at zero field were obtained, in other words,
no marked difference was observed in the present spectra below and above the metamagnetic transition at $B_{\rm M}$.
In contrast, with further increasing field above 1\,T, the quasielastic component shows remarkable broadening up to 5\,T.
The peak at 0.6\,meV is buried in the broadened quasielastic response with the field.
No additional peak was found under magnetic field.

Hereafter, we focus on the quasielastic response observed in {\ycz}.
The observed spectra can be reproduced by a sum of quasielastic Lorentzian and a Gaussian peak at 0.6\,meV or 2.1 meV.
The result was shown as solid lines in Fig.\ref{Fig:1}.
By fitting the spectra at 300\,mK at zero field, the quasielastic line width was derived to be 0.3\,meV.
This value coincides with the reported Kondo temperature of a few Kelvin, suggesting that the observed response arise from Kondo-type fluctuation.

This quasielastic component exhibits significant broadening with increasing field.
In Yb$T_2$Zn$_{20}$ system, strong reduction of heavy electron state was suggested in {\ycz} as well as YbIr$_2$Zn$_{20}$ based on specific heat, electrical resistivity and de Haas-van Alphen measurements\cite{Tak1,Hon,Ohy,Tak2}.
The ${\sqrt{A}}$-coefficient of the electrical resistivity in {\ycz} decreases two orders of magnitude decreases from zero field to 14\,T for the field along the $<$1\,0\,0$>$, which is supported by the effective mass derived from the dHvA measurements; the mass shows rapid decrease with increasing field.
The broadening of the quasielastic response with field in {\ycz} supports the reduction of the heavy-electron state from the microscopic point of view. 
Since significant anisotropic magnetic field response was found in {\ycz},
an experiment using single crystal is indispensable for further quantitative analysis.

\begin{figure}[t]
	\begin{center}
		\includegraphics[width=8cm]{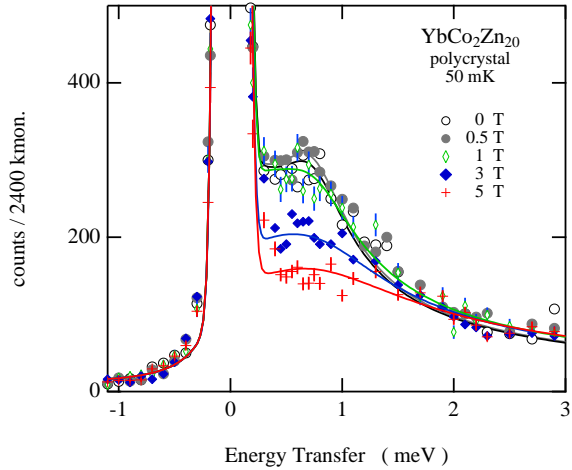}%
		\caption{\label{Fig:2}Magnetic field variation of inelastic scattering spectra of {\ycz} at 50\,mK under several magnetic fields up to 5\,T.}
	\end{center}
\end{figure}

Together with the quasielastic scattering, the distinct peaks were clearly observed, especially at zero field.
The observed peak can be ascribed to crystal-field excitation.
Takeuchi \textit{et al.} proposed a crystal-field level scheme for {\ycz} as ${{\Gamma}_6}$(0\,K) - ${{\Gamma}_8}$(9.1\,K) - ${{\Gamma}_7}$(28.5\,K) based on the results of specific heat, magnetic susceptibility and magnetization measurements\cite{Tak2} 
The peak at 0.6\,meV observed in the present study is consistent with  to the excitation from the ground state to the first excited state.
The second peak at 2.1\,meV and its evolution at high temperature of 10\,K agrees to that expected for the excitation from ${{\Gamma}_8}$ to ${{\Gamma}_7}$.
While there is small discrepancy in the energy of the first excited state, the overall splitting and its temperature dependence are consistent with the suggested model.

The present result clearly reveal the small overall splitting of the crystal field in {\ycz}.
On the other hand, the difference between the ground state and the first excited state is distinctly larger than the energy scale of Kondo temperature.  
This fact indicates that the large entropy of ${\sim}$8\,J/mol K$^2$ in {\ycz} does not come from degenerated crystal-field ground state, but mainly originate from the doublet ground state.

For more detailed discussion for anisotropy, magnetic correlation effects, and an origin of the field induced ordered phase, neutron scattering experiments using single crystals are in progress.

\section{Acknowledgments}
The authors would like to thank H. Yamauchi and Y. Shimojo for their assistance in the neutron scattering experiment. 
This work was supported by a Grant-in-Aid for Scientific Research on Innovative Areas "Heavy Electrons" (\#20102002, \#21102524, \#23102715),
on Priority Areas of New Material Science Using Regulated Nano Spaces (\#22013009, \#22013022),
and Young Scientists B (\#23740247) of The Ministry of Education, Culture, Sports, Science, and Technology, Japan.

\end{document}